\documentclass[11pt]{article}
\usepackage{amssymb}

\newcommand{\BE}{\begin{equation}}
\newcommand{\EE}{\end{equation}}
\newcommand{\BA}{\begin{eqnarray}}
\newcommand{\EA}{\end{eqnarray}}
\newcommand{\vol}{{\sf V}}

\begin{document}
\begin{titlepage}

\vspace*{1mm}
\begin{center}

            {\LARGE{\bf Vacuum condensates and `ether-drift' experiments}}

\vspace*{14mm}
{\Large  M. Consoli, A. Pagano and L. Pappalardo }
\vspace*{4mm}\\
{\large
Istituto Nazionale di Fisica Nucleare, Sezione di Catania \\
and Dipartimento di Fisica dell' Universit\`a \\
Via Santa Sofia 64, 95123 Catania, Italy}
\end{center}
\begin{center}
{\bf Abstract}
\end{center}
The idea of a `condensed' vacuum state is generally accepted
in modern elementary particle physics. We argue that 
this should motivate a new generation of precise
`ether-drift' experiments with present-day technology. 
\vskip 35 pt
\end{titlepage}


{\bf 1.}~The idea of a `condensed' vacuum is generally accepted
 in modern elementary particle
physics. Indeed, in many different contexts one introduces a set of 
elementary quanta whose perturbative `empty' vacuum state $|o\rangle$ is not
the physical ground of the interacting theory. 
In the physically relevant case of the Standard Model,
the situation can be summarized saying \cite{thooft} that 
 "What we experience as 
empty space is nothing but the configuration of the Higgs field that has the
lowest possible energy. If we move from field jargon to particle jargon, this 
means that empty space is actually filled with Higgs particles. They have 
Bose condensed." 

In this case, where the
condensing quanta are just neutral spinless particles 
(the `phions' \cite{mech}), 
the translation from `field jargon to particle jargon', 
amounts to establish a well defined functional relation (see ref.\cite{mech})
$n=n(\phi^2)$ between the average particle density 
$n$ in the ${\bf{k}}=0$ mode and the average value of the scalar 
field $\langle \Phi \rangle=\phi$. 
Thus, Bose condensation is just a consequence of
the minimization condition of the 
 effective potential $V_{\rm eff}(\phi)$. This has absolute
minima at some values $\phi =\pm v \neq 0$ for which $n(v^2)=\bar{n}\neq 0$
\cite{mech}. 

The symmetric phase, where $\langle \Phi \rangle=0$ and $n=0$, 
will eventually be re-established at a phase
transition temperature $T=T_c$. This, in the Standard Model, is so high 
that one can safely approximate the ordinary vacuum as a zero-temperature 
system (for comparison  think of $^4$He at a 
temperature of $10^{-12}$ $^o$K). 
 This observation provides the argument to
represent the vacuum as a quantum Bose liquid, i.e. a medium
where bodies can flow without any apparent friction, as in superfluid $^4$He,
in agreement with the experimental results.

On the other hand, the condensed particle-physics vacuum, while certainly different
from the ether of classical physics, is also different from
the `empty' space-time of Special Relativity which is assumed at the base of 
axiomatic quantum field theory.  
Therefore, following this line of thought, 
one may ask whether 
 the macroscopic occupation of the same quantum
state (${\bf{k}}=0$ in a given reference frame)
 can represent the operative construction
of a `quantum ether'  whose existence
might be detected through a
precise `ether-drift' experiment, of the type performed at the end of 
ninenteenth century and in the first half of twentieth century. This 
question leads
to the basic issue of a Lorentz-covariant description of the vacuum
that will be addressed in the following section.
\vskip 15 pt

{\bf 2.}~Although widely accepted, vacuum condensation is usually
considered just
a convenient way to rearrange the set of original degrees of freedom.
In this perspective, 
all differences between the physical vacuum and empty space are believed
to be reabsorbable into 
some basic parameters, such as the particle masses and few physical
constants, while leaving for the rest an exact Lorentz-covariant theory. 

On a more formal ground we observe, however, that the coexistence of 
{\it exact} Lorentz covariance and vacuum condensation in {\it effective}
quantum field theories is not so trivial. In fact, as a consequence of the
violations of locality at the energy scale fixed by the
ultraviolet cutoff $\Lambda$ \cite{salehi}, 
one may be faced with non-Lorentz-covariant 
{\it infrared} effects that depend on the vacuum structure.

This phenomenon can be understood in very simple terms starting from the
observation that, 
in a cutoff theory, the elementary
quanta are treated as `hard spheres' of radius $a \sim 1/\Lambda$,
as for the molecules of ordinary
matter. For the same reason, however, the simple idea
that deviations from Lorentz-covariance take only place at the cutoff scale
is incorrect. In fact, it is true that in the perturbative empty vacuum state 
(with no condensed quanta) non-locality is restricted to very short
wavelengths $2\pi/|{\bf{k}}|\leq a$. 
However, in a
condensed vacuum, the hard spheres will `touch' each other giving rise
to the propagation of {\it long-wavelength}
density fluctuations that, by definition, 
cannot be described in a Lorentz-covariant way.

To indicate this type of infrared-ultraviolet connection, originating 
from vacuum condensation in effective quantum field theories, 
Volovik \cite{volo1} has introduced a very appropriate name: 
reentrant violations of special relativity in the low-energy corner. 
In the simplest case of spontaneous symmetry breaking in a 
$\lambda\Phi^4$ theory, where the condensing
quanta are just neutral spinless particles, the `reentrant' effects reduce
to a small shell of three-momenta, say $|{\bf{k}}| < \delta$, where the 
energy spectrum deviates from a Lorentz-covariant form. Namely, 
by denoting $M_H$ as the typical energy scale associated with the 
Lorentz-covariant part of the energy spectrum, one finds
${{\delta}\over{M_H}} \to 0$ only when 
 ${{M_H}\over{\Lambda}}\to 0$. 

The basic ingredient to detect such `reentrant' 
effects in the broken phase consists in a purely
quantum-field-theoretical result:
the connected zero-four-momentum propagator $G^{-1}(k=0)$ 
is a two-valued function
\cite{legendre,pmu}. In fact, besides the well known solution
$G^{-1}_a(k=0)= M^2_H$, one also finds
$G^{-1}_b(k=0)= 0$. 

The b-type of solution
corresponds to processes where assorbing (or emitting)
a very small 3-momentum ${\bf{k}} \to 0$
does not cost a finite energy. This situation is well
known in a condensed medium, where a small 
3-momentum can be coherently distributed 
among a large number of elementary constituents, and corresponds to 
the hydrodynamical regime of density fluctuations whose
wavelengths $2\pi/|{\bf{k}}|$ are {\it larger} 
than $r_{\rm mfp}$, 
the mean free path for the elementary constituents. 

This interpretation \cite{weak,hierarchy}
 of the gap-less branch, which is very natural on the base of general 
arguments, is unavoidable in a superfluid medium. In fact, 
 "Any quantum liquid consisting of particles with integral 
spin (such as the liquid isotope $^4$He) must certainly have a spectrum of
this type...In a quantum Bose liquid, elementary excitations with small 
momenta ${\bf{k}}$  (wavelengths large compared with distances between atoms) 
correspond to ordinary hydrodynamic sound waves, i.e. are phonons. This 
means that the energy of such quasi-particles is a linear function of their
momentum" \cite{pita}. In this sense, a superfluid 
vacuum provides for ${\bf{k}} \to 0$ a universal picture.
This result does not depend
on the details of the short-distance interaction and even on the nature
of the elementary constituents. 
For instance, the same coarse-grained description is found in 
superfluid fermionic vacua \cite{volo2} that, as compared to the Higgs 
vacuum, bear the same relation of superfluid $^3$He to superfluid $^4$He.

Thus there are
two possible types of excitations with the same quantum numbers but
different energies when the 3-momentum ${\bf{k}} \to 0$: 
a single-particle
massive one, with ${E}_a({\bf{k}}) \to M_H$, and a collective
gap-less one with 
${E}_b({\bf{k}}) \to 0$.  `A priori', they can both propagate 
(and interfere) in the broken-symmetry phase.
Therefore, the situation is very similar to
superfluid $^4$He, 
where the observed energy spectrum is due to the peculiar
transition from the `phonon branch' to the `roton branch' at a momentum scale 
$|{\bf{k}}_o|$ where
\BE
{E}_{\rm phonon}({\bf{k}}_o) \sim 
{E}_{\rm roton}({\bf{k}}_o)
\EE
The analog for the Higgs condensate amounts to
an energy spectrum with the following limiting behaviours :

~~~i) ${E}({\bf{k}}) \to {E}_b({\bf{k}})= c_s |{\bf{k}}|$   
       ~~~~~~~~~~~~~~~~~~~~ for ${\bf{k}}\to 0 $

~~~ii) ${E}({\bf{k}}) \to {E}_a({\bf{k}}) = M_H+ {{ {\bf{k}}^2 }\over{2 M_H}}$
       ~~~~~~~~~~ for $|{\bf{k}}| \gtrsim \delta $

where the 
characteristic momentum scale $\delta \ll M_H$, at which
$E_a(\delta)\sim E_b(\delta)$, marks the transition from collective to 
single-particle excitations. This occurs for
\BE
\delta \sim 1/r_{\rm mfp} 
\EE
 where \cite{kine,seminar}
\BE
r_{\rm mfp} \sim {{1}\over{ \bar{n} a^2}}
\EE
is the phion mean free path, 
for a given value of the phion density $n=\bar{n}$ and a
given value of the phion-phion scattering length $a$. In terms of the 
same quantities, one also finds \cite{mech}
\BE
               M^2_H \sim \bar{n} a
\EE
giving the trend of the dimensionless ratios ($\Lambda \sim 1/a$)
\BE
\label{golden}
{{\delta}\over{M_H}} \sim {{M_H}\over{\Lambda}} \sim \sqrt{ \bar{n}a^3} \to 0
\EE
in the continuum limit where $a \to 0$ and the mass scale $\bar{n}a$ is
held fixed. 

By taking into account the above results, the physical
decomposition of the scalar field in 
the broken phase can be conveniently
expressed as (phys=`physical') \cite{physical}
\BE
\label{phime2}
\Phi_{\rm phys}(x) = v_R + {h}(x) + {H}(x)
\EE 
with 
\BE
\label{hh}
{h}(x)=
\sum_ { | {\bf {k}}| < \delta }  
\frac{1} { \sqrt{2 \vol {E}_k } } 
\left[  \tilde{h}_{\bf k}    {\rm e}^ { i ({\bf k}.{\bf x} -{E}_k t) } + 
(\tilde{h}_{\bf k})^{\dagger} {\rm e}^{-i ({\bf k}.{\bf x} -{E}_k t)} 
\right]
\EE
and
\BE
\label{H2}
{H}(x)=
\sum_{ |{\bf {k}}| > \delta }  
\frac{1} { \sqrt{2 \vol {E}_k } } 
\left[  \tilde{H}_{\bf k}    {\rm e}^ { i ({\bf k}.{\bf x} -{E}_k t) } + 
(\tilde{H}_{\bf k})^{\dagger} {\rm e}^{-i ({\bf k}.{\bf x} -{E}_k t)} 
\right]
\EE
where $\vol$ is the quantization volume and 
${E}_k=c_s|{\bf{k}}|$ for $|{\bf{k}}| < \delta$ while
${E}_k=\sqrt{{\bf{k}}^2 + M^2_H}$ for $|{\bf{k}}| > \delta$. Also, 
$c_s \delta \sim M_H$. 

Eqs.(\ref{phime2})-(\ref{H2}) replace
the more conventional relations
\BE
\label{conve1}
\Phi_{\rm phys}(x)= v_R + H(x)
\EE 
where
\BE
\label{conve2}
{H}(x)=
\sum_{ {\bf {k}} }
\frac{1} { \sqrt{2 \vol {E}_k } } 
\left[  \tilde{H}_{\bf k}    {\rm e}^ { i ({\bf k}.{\bf x} -{E}_k t) } + 
(\tilde{H}_{\bf k})^{\dagger} {\rm e}^{-i ({\bf k}.{\bf x} -{E}_k t)} 
\right]
\EE
with ${E}_k=\sqrt{{\bf{k}}^2 + M^2_H}$. 
Eqs.(\ref{conve1}) and (\ref{conve2})
are reobtained in the limit 
${{\delta}\over{M_H}} \sim {{M_H}\over{\Lambda}} \to 0$ 
where $h(x)$ disappears and the broken phase has
only massive excitations thus recovering an exactly Lorentz-covariant theory.
\vskip 15 pt
{\bf 3.}~Let us now return to the basic question posed at the end of 
Section 1. 
For finite values of $\Lambda$ there are long-wavelength density 
fluctuations of the vacuum and
Lorentz-covariance is not exact. Therefore, in the presence of such 
effects, can we try to 
detect the existence of the scalar condensate through a precise 
`ether-drift' experiment ? 

We first observe that
 a simple physical interpretation of the long-wavelength density 
fluctuation field 
\BE
\varphi(x)\equiv {{h(x)}\over{v_R}}
\EE
 has been proposed in 
refs.\cite{weak,hierarchy}. Introducing $G_F\equiv 1/v^2_R$ 
and choosing
the momentum scale $\delta$ as 
\BE
\label{choice}
                      \delta= \sqrt{  {{G_N M^2_H }\over{G_F}}  }
\EE
($G_N$ being the Newton constant) one obtains the identification
\BE
\varphi(x)=U_N(x)
\EE
$U_N(x)$ being the Newton potential. Indeed, with the choice in 
Eq.(\ref{choice}), to first order in $\varphi$ and 
in the limits of slow motions, the equations of
motion for $\varphi$ reduce to 
the Poisson equation for the Newton potential $U_N$ \cite{weak,hierarchy}
so that the deviations from
Lorentz covariance are of gravitational strength. If, as in the Standard Model,
$G_F$ is taken to be the Fermi constant one then
finds $\delta\sim 10^{-5}$ eV and $r_{\rm mfp}\sim 1/\delta=
{\cal O}(1)$ cm. As anticipated, the variation of
$\varphi(x)$ takes place over distances 
that are larger than $r_{\rm mfp}$ and thus
infinitely large on the elementary particle scale. Also, by introducing 
$M_{\rm Planck}= {{1}\over{ \sqrt {G_N} }}$, and using Eqs.(\ref{golden}) and
(\ref{choice}), one finds $\Lambda = q_H M_{\rm Planck}$ with 
$ q_H= \sqrt{ G_F M^2_H}= {\cal O}(1)$, or 
$a \sim 1/\Lambda \sim 10^{-33}$ cm.

At the same time, to first order, the observable 
effects of $\varphi$ can be re-absorbed \cite{hierarchy}
into an effective metric structure
\BE
ds^2= (1+ 2\varphi) dt^2 - (1-2\varphi)(dx^2 +dy^2 +dz^2)
\EE
that agrees with the first approximation to the line element of
 General Relativity \cite{rosen,weinberg}. 
In this perspective, 
the space-time curvature arises from a re-scaling
of the space-time units and from a refractive index for
light propagation 
\BE
\label{nphi}
                {\cal N}(\varphi)\sim  1- 2\varphi
\EE
so that the speed of light in the condensate frame 
(in units of $c=2.9979...10^{10}$ cm/sec) is
\BE
                     u\sim 1+ 2\varphi
\EE
Now, quite in general and
within Special Relativity (see page 145 of ref.\cite{synge}), 
a value ${\cal N}\neq 1$ implies a non-zero
drag coefficient $k$ 
\BE
              k=1- { {1}\over{ {\cal N}^2 }} \sim -4\varphi
\EE
so that, for an observer $S'$ moving with respect to
 the condensate frame with velocity $v$, and to first-order, 
light would propagate at a velocity
\BE
\label{pv}
                           u'(v)= u- kv
\EE
as for standard Galilei transformations with
a reduced relative velocity $kv$. 
\vskip 15 pt
{\bf 4.}~ This type of space-time picture leads naturally to 
the classical `ether-drift' experiments performed by
Michelson and Morley \cite{mm}, Illingworth \cite{illing} and
Miller \cite{miller} that have been
recently re-analyzed by Cahill and Kitto \cite{cahill}. 
Their conclusion is very simple and suggests the solution of
the long-standing problem concerning the 
nature of the observed effects. 
Namely, provided in those old 
experiments one takes into account the refractive index 
${\cal N}_{\rm medium}$ 
of the dielectric medium used in the interferometer
(air or helium), the observations
become consistent \cite{cahill} with the earth's velocity 
$v_{\rm earth} = 365\pm 18$ km/sec 
extracted from a fit to the COBE data for the
cosmic background radiation \cite{cobe}. In fact, the fringe shifts are
proportional to 
$  {{v^2_{\rm earth} }\over{c^2}} 
 (1- {{1}\over{ {\cal N}^2_{\rm medium} }})$ rather than to
                  ${{v^2_{\rm earth} }\over{c^2}}$ itself.

Cahill and Kitto used in their derivation a `Lorentzian' approach. In this
perspective, measuring devices are dynamically affected by their absolute
motion in such a way that this motion becomes unobservable 
\cite{bell,liberati}. However, if light propagates in a medium with 
${\cal N}_{\rm medium}\neq 1$, there is a small mismatch so that absolute 
motion may become observable.
In the following we shall argue that this effect is not in 
contradiction with Special Relativity. 

To this end, let us introduce an observer $S'$ that moves in an infinite, 
isotropical and homogeneous medium that defines an observer $S$. 
 Let us also consider 
two light beams, say 1 and 2, that are perpendicular in $S$ 
where they propagate along the $x$ and $y$ axis with velocities
$u_x(1)=u_y(2)=
u= {{c}\over{ {\cal N}_{\rm medium} }}$.  
Let us also assume that the velocity $v$ of  $S'$ is along the $x$ axis. 
 In this case, to evaluate the velocities of 1 and 2 for 
$S'$, we can apply Lorentz transformations with the result
\BE
 u'_x( 1)= {{ u -v }\over{ 1 - {{uv}\over{c^2}} }}~~~~~~~~~~~~~~u'_y(1)=0
\EE
and
\BE
 u'_x( 2)= -v~~~~~~~~~~~~~~u'_y(2)= u \sqrt{  1- {{v^2}\over{c^2}} }
\EE
In this way, a Lorentz transformation is equivalent to 
a local anisotropy that
becomes larger and larger by increasing the value of $v$.

Let us now define $L'_A$ and $L'_B$ to be the lengths of two 
optical paths, say A and B, as
measured in the $S'$ frame. For instance, they can represent the 
lengths 
of the arms of an interferometer which is at rest in the $S'$ frame. 
In the first experimental set-up, the arm 
of length $L'_A$ is taken along the direction of motion associated with the beam
1 while the arm of length $L'_B$ lies along the direction of the beam 2.
Notice that the two arms, in the $S'$ frame, form an angle
that differs from $90^o$ by ${\cal O}(v/c)$ terms.

Therefore, using the above results, 
the time for the beam 1 to go forth and back along $L'_A$ is 
\BE
          T'_A= L'_A (   
{{1- uv/c^2 }\over{u-v}} + {{1+ uv/c^2 }\over{u+ v}} )
\sim 
{{2L'_A}\over{u}} 
( 1+ k_{\rm medium} {{v^2}\over{u^2}} )
\EE
where
\BE
 k_{\rm medium}= 
1- {{1}\over{ {\cal N}^2_{\rm medium} }}
\EE
To evaluate the time $T'_B$, 
for the beam 2 to go forth and back along the arm of length
$L'_B$, one has first to compute
the modulus of its velocity in the $S'$ frame
\BE
  u'=  \sqrt{  
(u'_x(2))^2 + ((u'_y(2))^2 } = u \sqrt{ 1 + k_{\rm medium}{{ v^2}\over{u^2}} }
\EE
and then use the relation $u' T'_B= 2 L'_B$ thus obtaining
\BE
  T'_B = {{2 L'_B}\over{u'}} \sim 
{{2 L'_B}\over{u}}
( 1- k_{\rm medium} {{v^2}\over{2 u^2}} )
\EE
In this way, the interference pattern, between the light beam coming out 
of the optical path A and that coming out of the optical path B, is  
determined by the delay time
\BE
   \Delta T'= T'_A- T'_B \sim
{{2L'_A}\over{u}} 
( 1+ k_{\rm medium} {{v^2}\over{u^2}} )- 
{{2 L'_B}\over{u}}
( 1- k_{\rm medium} {{v^2}\over{2 u^2}} )
\EE
On the other hand, if the beam 2 were to propagate along the optical path
A and
the beam 1 along B, one would obtain a different 
delay time, namely
\BE
   (\Delta T')_{\rm rot}= (T'_A- T'_B)_{\rm rot}\sim 
{{2L'_A}\over{u}} 
( 1- k_{\rm medium} {{v^2}\over{2u^2}} )- 
{{2 L'_B}\over{u}}
( 1+ k_{\rm medium} {{v^2}\over{u^2}} )
\EE
so that, by rotating the apparatus, 
there will be a fringe shift proportional to 
\BE
\label{deltat}
(\Delta T')- (\Delta T')_{\rm rot} \sim 
{{3(L'_A+ L'_B)}\over{u}}  k_{\rm medium} {{v^2}\over{u^2}}
\EE
In this way $S'$ will now be able to determine its `absolute'
velocity in complete agreement with Special Relativity. In fact, for $S'$
Eq.(\ref{deltat}) is the only way to detect the existence of the 
$S$ observer through 
the value of a velocity $v$ whose operative definition, otherwise, 
would be unclear (dealing with a uniform motion
in an infinite, isotropical and homogeneous medium).
 On the other hand, if the numerical value 
${\cal N}_{\rm medium} \neq 1$ were unknown, $S'$ would try to determine $S$
through an effective , reduced velocity 
$v_{\rm obs} \sim \sqrt {k_{\rm medium} } v$ rather than through $v$ itself.

Now, the following question naturally arises. What happens if we 
remove the medium everywhere except in a small region of space 
that includes the arms of the interferometer ?
Will the fringes shift upon rotation of the apparatus ? At first sight, the
answer is positive. In fact, the occurrence of a fringe shift by rotating
an apparatus at rest in the $S'$ frame
cannot depend on the presence of the medium in the outer 
regions of space. After some thought, however, the answer might 
become negative. Indeed, one may argue that 
the medium is now taken at rest in
the $S'$ frame so that the two light beams should propagate with the
same velocity, regardless of their orientation. 

The latter expectation is based on considering now
the observer $S'$ to be physically equivalent to 
the observer $S$ introduced before (for which we assumed
an exactly isotropical value 
$u= {{c}\over{ {\cal N}_{\rm medium} }}$ everywhere). However, 
this equivalence has no rigorous basis since, differently from 
$S'$, the observer $S$ was taken at rest in an {\it infinite} medium. 

Therefore, the occurrence (or not) of fringe shifts becomes a purely
experimental issue \cite{notesynge}:
 a way to {\it test} local isotropy. 
In practice, for the earth's velocity, and to
 ${\cal O}({{v^2_{\rm earth} }\over{c^2}} )$, one can re-analyze 
\cite{cahill}
the experiments in terms of the effective parameter
\BE
                \epsilon = {{v^2_{\rm earth} }\over{u^2}} k_{\rm medium}
\equiv {{v^2_{\rm obs} }\over{c^2}}
\EE
and use the relevant experimental values
${\cal N}_{\rm air}\sim 1.00029$ or
${\cal N}_{\rm helium}\sim 1.000036$.  

For instance, for
$v_{\rm earth}=365\pm 18 $ km/sec (and an in-air-operating optical system)
one predicts 
    $\epsilon\sim 10^{-9}$ or
$v_{\rm obs} \sim 9 $ km/sec, precisely 
Miller's result. 

The comparison with the experiment of Kennedy and 
Thorndike can also be done along similar lines
 by restricting their analysis to
long-period observations where they found
 a non-zero value $v_{\rm obs}= 15 \pm 4 $ km/sec \cite{kt}.

Notice that the same analysis might even be applied to the
experiment by Jaseja {\it et al} 
\cite{jaseja} where one was measuring the
shift $\Delta \nu$ in the maser frequency $\nu_c$
introduced by the rotation of the apparatus so that
${{\Delta \nu}\over{\nu_c}} \sim \epsilon$. 
Indeed, the results were showing a well defined shift 
$\Delta \nu\sim$275 kHz or roughly one part over $10^9$ in the basic
frequency $\nu_c\sim 3\cdot 10^{14}$ Hz. However, 
this experimental effect, well consistent with Miller's results \cite{note2},
 was not taken seriously and considered to be spurious 
"..presumably due to magnetostriction in the Invar spacers due to the
earth's magnetic field". To obtain a consistency check of this
interpretation, the authors of ref.\cite{jaseja}
were indeed planning to repeat their analysis by replacing the potentially 
problematic parts of their apparatus. However, this improved experiment
was never performed \cite{selleri}.
\vskip 15 pt

{\bf 5.}~We are now ready to return to 
the density fluctuations of the scalar condensate discussed in sect.3. 
 To this end we observe
that, according to Cahill and Kitto \cite{cahill} (and according to our
previous analysis) in the
vacuum experiments of Joos \cite{joos} and
of Brillet and Hall \cite{brillet} no effect 
could have been observed. In fact, in this case, 
${\cal N}_{\rm vacuum}=1$ exactly so that $v_{\rm obs}=0$.
 However, even the very precise Brillet and Hall 
experiment might be considered 
as showing a non-zero result, although at a level of
accuracy $\sim 10^{-15}$ (see their figure 3 and the associated 
figure caption). Thus one might
speculate on the possible effect of the non-zero refractive index 
Eq.(\ref{nphi}) for which, `in the vacuum', there should be, nevertheless, 
corrections proportional to 
\BE
    {{v^2_{\rm earth} }\over{c^2}} 
(1- {{1}\over { {\cal N}^2(\varphi) }}) 
\EE
Now, transforming to ordinary units, and for a centrally symmetric
field, one has
\BE
            \varphi (R) =- {{G_N M}\over{c^2 R}}
\EE
Therefore, for an apparatus placed on the earth's surface, one finds
$\varphi(R)\sim -0.7\cdot 10^{-9}$ (for $M=M_{\rm earth}$ and 
$R=R_{\rm earth}$) and
\BE
{{v^2_{\rm earth} }\over{c^2}} 
(1- {{1}\over { {\cal N}^2(\varphi) }}) 
\sim 4\cdot 10^{-15} 
\EE
for the same value $v_{\rm earth} = 365\pm 18$ km/sec extracted
from the COBE data. This tiny effect, which 
is consistent with the results obtained by
Brillet and Hall, might be detected in a more precise 
experiment, with present-day technology and
level of accuracy $\sim 10^{-16}$. 

Summarizing: according to current ideas, the vacuum is not `empty'.
Thus, one should carefully check the compatibility between 
exact Lorentz
covariance and vacuum condensation in effective
quantum field theories. For the specific case of the scalar condensate,
 the non-locality
associated with the presence of the ultraviolet cutoff will also
show up at long wavelengths in the form of non-Lorentz-covariant 
density fluctuations 
associated with a scalar function $\varphi(x)$. 

If, on the base of refs.\cite{weak,hierarchy}, these long-wavelength
effects are naturally interpreted in terms of the Newton potential $U_N$
(with the identification $\varphi=U_N$), one obtains
the weak-field space-time curvature of General Relativity
and a refractive index ${\cal N}\sim 1-2\varphi$. This value of 
${\cal N}(\varphi)$ might be important to understand a very 
precise (vacuum) `ether-drift' experiment with present-day technology
and level of accuracy $\sim 10^{-16}$. 
\vskip 30 pt
{\bf Acknowledgements}
\vskip 5 pt
We thank D. Maccarrone and F. Selleri for useful discussions.


\begin{thebibliography} {99}
\bibitem{physical}
By `physical' we mean that the normalization of the vacuum field
$v_R$ is such that 
the quadratic shape of $V_{\rm eff}(\phi_R)$ at $\phi_R=v_R$ is precisely 
given by the physical Higgs mass $M^2_H$.
\bibitem{notesynge}
 In this sense, as already
remarked by Synge (see page 161 of ref.\cite{synge}), it is hard to make
definite predictions about `ether-drift' 
experiments on the base only of Special Relativity.
\end{thebibliography}
\end{document}